\documentstyle[12pt,aaspp4]{article} 
\def\tempest%
{\begin{array}{ccc} 
1 & 1 & 1 \\ 
1 & 1 & 1 \\ 
4 & 3 & 8 
\end{array}}

\def\kms{{\rm km}\,{\rm s}^{-1}} 
 
\def\tot{{\rm tot}} 
\def\orb{{\rm orb}} 

\begin{document}

\title{Binary Black Hole Mergers from Planet-like Migrations}
\author 
{Andrew Gould}
\affil{Ohio State University, Department of Astronomy, Columbus, OH, USA} 
\affil{E-mail: gould@astronomy.ohio-state.edu} 
\author 
{Hans-Walter Rix}
\affil{Max-Planck-Institut f\"ur Astronomie, Heidelberg, Germany}
\affil{E-mail: rix@mpia-hd.mpe.de} 

\begin{abstract} 
If supermassive black holes (BHs)
 are generically present in galaxy centers, and if 
galaxies are built up through hierarchical merging, BH binaries
are at least temporary features of most galactic bulges. Observations suggest,
however, that binary BHs are rare, pointing towards a binary lifetime
far shorter than the Hubble time. We show that, regardless of the detailed
mechanism,  all stellar-dynamical processes
are insufficient to reduce significantly the orbital separation once 
orbital velocities in the binary exceed the virial velocity of the system.
We propose that a massive gas disk surrounding a BH binary can effect its
merger rapidly, in a scenario analogous to the orbital decay of super-jovian
planets due to a proto-planetary disk.  As in the case of planets, gas
accretion onto the secondary (here a supermassive BH) is integrally connected
with its inward migration.  Such accretion would give rise to quasar activity.
BH binary mergers could therefore be responsible for many or most quasars.

\end{abstract}

\section{Introduction} 

        Supermassive black holes (BH) are nearly ubiquitous in nearby galaxy
nuclei (e.g.\ Ho 1999).  These BHs formed very early, probably
during the epoch of quasars, $z\ga 2$, and are now largely dormant
remnants of quasars.  In the hierarchical
picture of structure formation, present day galaxies are
the product of successive mergers (e.g.\ White 1996), 
and indeed there is evidence for many mergers 
in the high-$z$ universe (Abraham et al.\ 1996).  
Hence, it appears almost inevitable that
modern galaxies should harbor, or at least should have once harbored, 
multiple BHs that were collected during their merger history
(Kauffman \& Haehnelt 1999).

        BHs of mass $M\ga 10^7\,M_\odot$ will quickly find their
way to the center of a merger remnant by dynamical
friction.  Logically, there are only three possibilities.  First,
BH pairs could merge to form a single, larger BH.  Second,
the pairs of BHs could form binaries that would remain at galaxy centers
to this day.  Finally, a third BH could also fall in,
leading to a three-body interaction violent enough to expel any number of
the three BHs from the galaxy
 (Begelman, Blandford, \& Rees 1980).  While in principle
this means that all three holes could be ejected, in practice such a violent
ejection event is unlikely unless the binary's internal velocity
is much higher than the escape velocity from the galaxy ($\ga 2000\,\kms$);
in this case, the binary would be in the late stages of merging
anyway (see \S\ 2).  Since
the broad lines of quasars are not often observed to be displaced from the
narrow lines by such high velocities, the fraction of binaries with such
high internal velocities cannot be large, and therefore triple ejection cannot
be common.  Hence, mergers generically produce BH binaries, and these
binaries either merge on timescales short compared to a Hubble time, or they
are present in galaxies today.

        Observationally, there is evidence only for a few 
massive BH binaries (e.g., Lehto \& Valtonen 1996)
and in none of these cases is the evidence
absolutely compelling.  Theoretically, it has proven difficult to
construct viable merger scenarios for these BH binaries.  Here we first
review this difficulty of driving the merger by the 
stellar-dynamical means that are discussed in the literature.
We then propose a gas-dynamical alternative.

\section{Near Impossibility of Stellar Dymnamics-Driven Mergers}

        If a BH binary could (somehow) be driven to a sufficiently
small orbit, then gravitational radiation would increasingly sap energy
from the system and so engender a merger.  For a circular orbit with an
 initial velocity $v_{\rm gr}$, the time $T$ to a merger due
 to gravitational radiation is given by 
\begin{equation}
v_{\rm gr} = c \biggl({5\over 256}\,{G M_\tot^2\over \mu T c^3}\biggr)^{1/8}
= 3400\,\kms\,\biggl({M_\tot^2/\mu\over 8\times 10^8\,M_\odot}\biggr)^{1/8}
\biggl({T\over 10\, \rm Gyr}\biggr)^{-1/8}
\label{eqn:gravrad}
\end{equation}
where $M_\tot=M_1+M_2$ is the total mass,
$\mu=M_1 M_2/M_\tot$ is the reduced mass, and where we have normalized
to the case $M_1=M_2=10^8\,M_\odot$.
Note that for fixed total mass, the equal-mass case gives a lower limit
on this required velocity, and that the result depends only very weakly
on the total mass.

        However, as we now show it is almost impossible to
achieve this velocity by any conceivable stellar-dynamical process.
The basic problem is that when the orbital velocity $v_\orb$ 
is about equal to the stellar velocity
dispersion $\sigma\sim 200\,\kms$,
the total mass in stars within a volume circumscribed by the BH
orbital radius $(a\sim 5\,{\rm pc}\,M_\tot/10^8\,M_\odot)$ 
is about $M_\tot$.  If all of these stars were expelled from the
BH binary at speed $v_\orb(M_2/M)^{1/2}$ 
(Rajagopal \& Romani 1995 and references therein)
the binding
energy of the binary would increase by only a factor $\sim e$.  
However, to get from a virial velocity of $\sim 200$km/s to
$v_{\rm gr}$ (eq.\ \ref{eqn:gravrad}), would require $N_e\sim 6$ $e$-foldings 
in binding energy.
Hence, the binary will clear out a hole in the stellar distribution, and 
dynamical friction will be shut down 
(Quinlan 1996; Quinlan \& Hernquist 1997).

        The most efficient conceivable process to rejuvenate the orbital
decay would be to equip the binary with an intelligent ``captain''.  
Like a fisherman working in over-fished waters, whenever the captain saw 
that the binary was running out of stars to expel, she would steer the binary
to the densest unexploited region of the galaxy.  To effect the merger,
this would mean systematically moving 
through and expelling all the stars within a region containing about
$N_e\,M_\tot$ in stars.  For a galaxy with an $r^{-2}$ density
profile, this implies expelling all the stars within a radius
$r=N_e G M_\tot/2\sigma^2\sim 60\,$pc, where we have made the evaluation for
$M_\tot=2\times 10^8\,M_\odot$ and $\sigma=200\,\kms$.

        The real difficulty of the captain's work is best understood by
considering the last $e$-folding before gravitational radiation can take over.
For $v_\orb\gg\sigma$, the cross section for hard interactions 
(including gravitational focusing) is $\la \pi a^2 v_\orb/\sigma$.
If each incident particle is expelled with speed $v_\orb(M_2/M)^{1/2}$
(Rajagopal \& Romani 1995), then the binding
energy $E_b$ decays at $d\ln E_b/d t \sim 2\pi a^2v_\orb\rho /M
=G\rho P$, where $P$ is the period, and
$\rho$ is the local density.  The last $e$-folding alone would require a time
$t\sim [G\rho(r)P]^{-1}\sim 
2\pi (r/\sigma)^2/P\sim
2\,$Gyr, where
we have assumed $r\sim 30\,$pc and our other canonical parameters.  Thus,
even with the captain's careful guidance, the full merger requires a large
fraction of a Hubble time.  Moreover, comparing this decay rate
with the standard formula
for the decay of translational energy $E_t$ (Binney \& Tremaine 1987)
yields,
\begin{equation}
{d \ln E_b\over d t} \la 0.1\,\biggl({\sigma\over v_\orb}\biggr)^3 
{d \ln E_t\over d t}.
\label{eqn:dynfric}
\end{equation}
That is, $d\ln E_b/d\ln E_t \la 10^{-4}$, so that
the binary would be driven by dynamical friction
back to the center of the Galaxy before it
had completed $10^{-4}$ of an $e$-folding of energy loss.  Hence,
the captain would have to initiate $10^4$ ``course changes'' in the last
$e$-folding alone.  Since the ``captain'' must in fact be some random
process, the only source of such ``course changes'' is
brownian motion due to continuous interaction with other compact
objects. However, for stars of mass $m$ in an $r^{-2}$ profile, the
range of such Brownian motion is $\Delta\ln r\sim m/M_\tot$, i.e., too 
small by several orders of magnitude.  In contrast to ordinary Brownian
motion, the present system has an ``external'' energy source, the binary's
binding energy.  However, it follows from equation (\ref{eqn:dynfric})
that even if all of this donated energy were acquired by the binary's 
transverse motion, the brownian motion would be only slightly augmented.
In any event, most of the donated energy goes to the stars, not the binary.
Infall of globular clusters might well give the binary an occassional jolt,
but these would be far too infrequent to drive the merger.  
In brief, any sort of mechanism to drive a merger by
ordinary dynamical friction, no matter how contrived, is virtually ruled out.

        The only loophole to this argument is that we have assumed
circular binary orbits.  If an instability existed that systematically
drove the BH binaries toward eccentricity $e\rightarrow 1$
orbits, then either the binaries would suffer enhanced gravitational
radiation (for a fixed semi-major axis) or could even
merge in a head-on collision.  Fukushige, Ebisuzaki, \& Makino (1992)
 first suggested such an
instability based on the following qualitative argument: dynamical
friction is more effective at low speeds than high speeds and hence,
in the regime where the ambient particles interact with the binary mainly
by encounters with its individual members $(v_\orb\la\sigma)$, 
the binary would suffer more
drag at apocenter than pericenter, tending to make the orbit more eccentric.
Fukushige et al.\ (1992) 
presented numerical simulations that gave initial support to
this conjecture.  There are, however, two reasons for believing that this
effect cannot drive mergers.  First, several groups have conducted more
sophisticated simulations, and these do not show any strong tendency for
$e\rightarrow 1$ (Makino et al.\ 1994; Rajagopal \& Romani 1995;
Quinlan \& Hernquist 1997).  Second, once the binary entered the regime
$v_\orb\gg \sigma$, the ambient particles would interact with the binary as
whole, and so there is no reason to expect any drive toward high 
eccentricities.  Hence, while this loophole is not definitively closed,
neither does it look particularly promising.

\section{Gas Dynamical Solution}

Begelman, Blandford \& Rees (1980) were the first to suggest 
that gas infall may ``lead to some orbital evolution''. But, at the time
it was not clear that all other mechanism to overcome the BH hangup
would most likely fail. 

 To resolve the above dilemma, we suggest that gas dynamics play the decisive
role in orbital decay, forcing the secondary BH  to ``migrate''
in toward the primary in a manner analogous to the migration of planets.
Such migration has been proposed to account for the discovery of jovian-mass
and superjovian-mass planets at $\la 1\,$AU from solar-type stars,
while it is generally believed that such massive
planets can only be created several AU from the stars (Trilling et al.\ 1998).
Artymowicz \& Lubow (1994, 1996) simulated 
interactions between moderately unequal-mass binaries
and accretion disks, which is more directly relevant to the present case
than extreme-ratio (planetary) systems.  They did not follow the 
orbital evolution as has been done in more recent work on planets, but
only evaluated the instantaneous effect of the torques.  They found 
a migration to higher eccentricities was a larger effect than migration
to smaller orbits.
Regardless of which effect dominates, one
would expect the final merger to be from circular rather than radial
orbits: if the binary is driven toward radial orbits, its emission of
gravitational radiation near pericenter will eventually pull in the
apocenter of the orbit, decoupling the binary from the disk and allowing
the gravitational radiation to circularize the orbit before final
coalescence.

For migration to work, the galaxy merger that creates the BH binary
must eventually dump at least $M_2$ worth of gas into the inner 
$\sim 5\,$pc of the merger remnant where the binary coalescence has gotten
``hung up''.  Whether this happens on timescales short compared to a dynamical
time at 5 pc ($\sim 10^7\,$yr), leading to tremendous gas
densities and ensuing rapid star formation (Taniguchi \& Wada 1996),
 or whether the gas accumulates 
over a longer timescale and so does not trigger a starburst, the
basic scenario will be the same.
  
   There is every reason to expect mergers effect such a gas accumulation.
 First, quasars must gorge themselves on gas to reach
their present size.  Hence, regardless of whether our picture of binary
mergers is correct,  this much gas must find its way to central BHs.
Second, there is substantial evidence that many quasars
are in either
recent merger remnants or at least significantly disturbed galaxies
(Kirhakos et al.\ 1999 and references therein).  
Hence, it seems likely that mergers are the most efficient
means to drive gas to the center.  Third, many spiral bulges and ellipticals
have cuspy profiles populated by metal rich stars whose total mass
is comparable to that of their massive BHs (van der Marel 1999).  Thus, it must
be possible to funnel huge amounts of gas to the centers of galaxies.

        In planet migration, the migration timescale is similar to the
accretion timescale for growing the planet because the two processes
are governed by the same phenomena, gravitational torques and dissipation
(Trilling et al.\ 1998; A.\ Nelson 1998, private communication).  
We expect the same to be true of migration of BH binaries.
Thus, there should be a grand accretion disk around the primary with
a ``gap'' opened up by the secondary.  Material should be transported
across this gap to a second, smaller accretion disk surrounding the secondary
BH.  
The total energy liberated by this smaller accretion disk should be
$\sim \epsilon M_2 c^2 \sim 2\times 10^{61} (M_2/10^8 M_\odot)\,$ergs,
where we have taken the efficiency to be $\epsilon=0.1$, producing a
quasar-like appearance during this phase.

\section{Discussion}

        While our suggestion, driven by the lack of alternatives,
makes few unambigous predictions, it does
open several lines of investigation that could help test and flesh out
our picture.

        First, merging binaries would appear very much like
quasars, since our picture of the migrating secondary is essentially
identical to the standard picture of a quasar.  The one difference is
that the jet from a migrating BH could precess if the orbit of
the secondary were substantially misaligned relative to the accretion disk.
At present, however, we have no method of estimating how often 
significant misalignment should occur.

        Second, the redshift of the broad lines from a migrating BH's
accretion disk should be offset from the redshift of the host galaxy (as
traced perhaps by the narrow lines).  Since the migration probably
accelerates with time, most migrating quasars should have $v_\orb\sim \sigma$.
Nevertheless, some should have substantially higher offsets, and measuring
the distribution of these offsets would allow one to trace the
migration process.  However, if no offsets were observed, this would not
in itself rule out our hypothesis.  It could be, for example, that
migrating binaries in merger remnants are preferentially buried in a
larger, roughly spherical cloud of dusty gas.  In this case, they would
have more similarity to ultra-luminous infrared galaxies (ULIRGs)
than to quasars, and
the line centers of their emission would be at the galaxy velocity, not
that of the secondary.

        Third, it is at least possible that one would see two broad-line 
systems, one from the primary and one from the secondary.  Since broad
lines are by definition broad ($\ga 3000\,\kms$), the existence of two
systems would not easily be recognized for $v_\orb\sim\sigma$.  
However, distinct
peaks might be discernible when the BHs were closer to merger.
On the other hand, it may be that the major supply of gas lies outside
the orbit of the secondary, and hence the primary does not generate
a significant broad-line region.

        Fourth, it will be important to carry out simulations to determine
whether the gas dissipation timescale is short enough for the acceting
material to follow the binary inward.  This is certainly the case for
the simulations that have been done for extreme mass-ratio (planetary)
systems, but needs to be checked for the less extreme case also.

        Finally, we suggest that migrating BH binaries may
simply {\it be} the quasars, or at least most of them.  They have
the same integrated energy output as quasars, they have the same
accretion-disk fuel source as quasars, and like quasars, they turn on
in the wake of mergers.  It may be easier to move gas inward from 
$\sim 5\,$pc scales for a binary BH than for a single BH because the
binary would excite spiral density waves in the grand accretion disk
and so augment viscous drag.  Accretion in the inner disk around the
secondary might also be easier than for an isolated BH because
of the tidal effects of the primary.  If this hypothesis
is correct, then quasars should generically show offsets between the centers
of their broad and narrow lines with a root mean square of 
$\sim (2/3)^{1/2}\sigma$.


\bigskip 

{\bf Acknowledgements}: 
We thank Andy Nelson for valuable discussions.
A.G.\ thanks the Max-Planck-Institut f\"ur Astronomie for its hospitality
during a visit when most of work of this Letter was completed.
His work was supported in part by grant AST 97-27520 from the NSF.


\end{document}